# Do we live in the universe successively dominated by matter and antimatter?

Dragan Slavkov Hajdukovic[1]
PH Division CERN
CH-1211 Geneva 23
dragan.hajdukovic@cern.ch
[1]On leave from Cetinje, Montenegro

**Abstract**

We wonder if a cyclic universe may be dominated alternatively by matter and antimatter. Such a scenario demands a mechanism for transformation of matter to antimatter (or antimatter to matter) during the final stage of a big crunch. By giving an example, we have shown that in principle such a mechanism is possible. Our mechanism is based on a hypothetical repulsion between matter and antimatter, existing at least deep inside the horizon of a black hole. When universe is reduced to a supermassive black hole of a small size, a very strong field of the conjectured force might create (through a Schwinger type mechanism) particle-antiparticle pairs from the quantum vacuum. The amount of antimatter created from the vacuum is equal to the decrease of mass of the black hole and violently repelled from it. When the size of the black hole is sufficiently small, the creation of antimatter may become so fast, that matter of our Universe might be transformed to antimatter in a fraction of second. Such a fast conversion of matter into antimatter may look as a Big Bang. Our mechanism prevents a singularity; a new cycle might start with an initial size more than 30 orders of magnitude greater than the Planck length, suggesting that there is no need for inflationary scenario in Cosmology. In addition, there is no need to invoke CP violation for explanation of matter-antimatter asymmetry. Simply, our present day Universe is dominated by matter, because the previous universe was dominated by antimatter.

The idea of antigravity (defined as the gravitational repulsion between matter and antimatter) is as old as the discovery of antimatter (For a review see Nieto and Goldman, 1991). In the early sixties of the 20[th] century, antigravity was abandoned by main-stream physics, not because of experimental evidence against it, but because of theoretical arguments (Morrison, 1958; Schiff, 1958, 1959; Good, 1961) believed to be out of any reasonable doubt. While opposing the idea of antigravity, the paper of Nieto and Goldman (1991) contains a critical reconsideration of the old arguments leading to conclusion that they are still sufficiently strong to exclude antigravity but not without shortcomings; in the light of the new knowledge the arguments were less convincing in nineties than in sixties. The arguments against antigravity were further questioned by Chardin and Rax (Chardin and Rax, 1992; Chardin, 1993, 1997) with intriguing arguments that CP violation might be a consequence of antigravity and a recent paper by Villata (2011) arguing that "antigravity appears as a prediction of general relativity when CPT is applied". Additionally, assuming the existence of antigravity, Hajdukovic (2007, 2010a, 2010b, 2010c and 2011) has considered phenomena related to the gravitational version of the Schwinger's mechanism (Schwinger, 1951) and the gravitational polarization of the quantum vacuum. Hence, after nearly half a century of suppression, the idea of antigravity is back. Of course, if antigravity exists or not, can be revealed only by the future experiments, like the AEGIS experiment (Kellerbauer A et al. 2008) at CERN designed to measure the gravitational acceleration of anti-hydrogen. Complementary information might come from the neutrino astronomy if, as predicted recently (Hajdukovic, 2007), the supermassive black holes behave as point-like sources of antineutrinos.

In the present Letter, as an illustration of possible consequences of antigravity, we point out a new scenario for a cyclic universe.

Soon after Einstein's foundation of the General Relativity it was understood (Friedman, 1922) that it is compatible with the idea of a cyclic universe. In the framework of these first models the question was if the matter-energy density in the Universe is sufficiently large (i.e. larger than a critical value $3H^2/8\pi G$) to provoke a future collapse of the Universe ending with a Big Crunch, eventually followed by a new Big Bang.

The 21st century has started with a proliferation of much more sophisticated cyclic models in the framework of different theories like quantum loop gravity, braneworld models, conformal cosmology and so on (see review by Novello and Bergliaffa, 2008). An inspection of the existing models shows that that in spite of great differences between them they have a common point: all cycles are dominated by matter.

In the present Letter, contrary to all previous models, we present a radically new possibility that we live in a cyclic universe dominated alternatively by matter and antimatter; a universe dominated by matter (as it is the universe in which we live) is always followed by a universe dominated by antimatter and vice versa.

The aim of the Letter is modest. We do not develop a new cyclic model of the universe; we have only proposed a mechanism allowing transition from a matter to an antimatter universe and vice versa. The further development may go in two directions: detailed study of the proposed mechanism or the discovery of alternative mechanisms which might produce the same phenomenon.

Without entering complex discussions, the simplest way to postulate a gravitational repulsion between matter and antimatter is

$$m_i = m_g \; ; \; m_i = \overline{m}_i \; ; \; m_g + \overline{m}_g = 0, \qquad (1)$$

where a symbol with a bar denotes antiparticles; while indices $i$ and $g$ refer to the inertial and gravitational mass (gravitational charge). The first two relations in (1) are in the same time the experimental evidence (Will, 1993; Gabrielse, 1999) and the cornerstone of the General Relativity; while the third one is the conjecture of antigravity which dramatically differs from the mainstream conviction $m_g - \overline{m}_g = 0$, implying (together with the Newton law of gravity) that matter and antimatter are mutually repulsive but self-attractive. In simple words, while an apple falls down, an anti-apple would fall up.

As an alternative to the above long-range antigravity we can imagine existence of a matter-antimatter repulsion (of gravitational or non-gravitational origin) which is significant only deep inside the horizon of a black hole; hence, the range of interaction is much smaller than the Schwarzschild radius.

Let us consider the simplest case of a Schwarzschild black hole made from matter. While it is often neglected, from mathematical point of view there are two solutions: the positive mass Schwarzschild solution

$$ds^2 = c^2\left(1 - \frac{2GM}{c^2 r}\right)dt^2 - \left(1 - \frac{2GM}{c^2 r}\right)^{-1} dr^2 - r^2 d\theta^2 - r^2 \sin^2\theta \, d\phi^2, \qquad (2)$$

considered as the physical space-time metric; and the negative mass Schwarzschild solution

$$ds^2 = c^2\left(1 + \frac{2GM}{c^2 r}\right)dt^2 - \left(1 + \frac{2GM}{c^2 r}\right)^{-1} dr^2 - r^2 d\theta^2 - r^2 \sin^2\theta \, d\phi^2, \qquad (3)$$

considered as a nonphysical solution. It serves as the simplest example of a naked singularity (Preti and Felice, 2008; Luongo and Quevedo, 2010) and a repulsive space-time allowed by mathematical structure of general relativity but rejected as nonphysical. However, in the framework of the gravitational repulsion between matter and antimatter, both solutions may be given a physical meaning: the metric (2) is metric "seen" by a test particle, while the metric (3) is metric "seen" by a test antiparticle.

The major difference is that there is a horizon in the case of metric (2), while there is no horizon in the case of metric (3). In simple words, a black hole made from matter, acts as a black hole with respect to matter and, as a white hole with respect to antimatter.

According to the metric (3) the radial motion of a massive antiparticle is determined by

$$\dot{r}^2 = c^2(k^2 - 1) - \frac{2GM}{r}, \qquad (4)$$

where $k$ is a constant of motion and dot indicates derivative with respect to the proper time.

Differentiating the equation (4) with respect to proper time and dividing through $\dot{r}$ gives

$$\ddot{r} = \frac{GM}{r^2}. \qquad (5)$$

The equation (5) has the same form as should have the corresponding Newtonian equation of motion with the assumed gravitational repulsion.

Of course, in spite of the same form of Newtonian equations and the equations (5) coming from general relativity, there is fundamental differences between them. The coordinate $r$ in equation (5) is not the radial distance as it is in the Newtonian theory, and dots indicate derivatives with respect to proper time, rather than universal time.

In order to understand the physical significance of the conjecture (1), we must remember the Schwinger mechanism (Schwinger, 1951) in Quantum Electrodynamics: a strong electric field $E$, greater than a critical value $E_{cr}$, can create electron-positron pairs from the quantum vacuum. For instance, electron-positron pairs can be created in the vicinity of an artificial nucleus with more than 173 protons (Greiner et al., 1985; Ruffini at al., 2010).

In the case of an external (classical i.e. unquantized) constant and homogenous electric field $E$ the exact particle creation rate per unit volume and time is

$$\frac{dN_{m\bar{m}}}{dtdV} = \frac{4}{\pi^2} \frac{c}{\lambdabar_m^4} \left(\frac{g}{g_{cr}}\right)^2 \sum_{n=1}^{\infty} \frac{1}{n^2} \exp\left(-\frac{n\pi}{2} \frac{g_{cr}}{g}\right); \quad g_{cr}(m) = \frac{2c^2}{\lambdabar_m}, \qquad (6)$$

where $\lambdabar_m \equiv \hbar/mc$ denotes the reduced Compton wavelength corresponding to the particle with mass $m$. Let us observe that we have replaced the quotient of electric fields $E/E_{cr}$ (appearing in Quantum Electrodynamics) by the quotient of corresponding accelerations $g/g_{cr}$; so that the result (6) could be used not only in the case of an electric field, but also in the case of antigravity.

If $g > g_{cr}$, the infinite sum in Eq. (6) has numerical value not too much different from 1. So, a simple, but good approximation is:

$$\frac{dN_m}{dtdV} \approx \frac{4}{\pi^2} \frac{c}{\lambdabar_m^4} \left(\frac{g}{g_{cr}(m)}\right)^2. \qquad (7)$$

The Schwinger mechanism has two cornerstones, the first one is the existence of quantum vacuum and the second one the existence of an external electric field (which attempts to separate electrons and positrons).

Before the foundation of quantum field theory (QFT), the physical vacuum was synonym for nothing. However, in quantum field theory "nothing's plenty" as nicely said by Aitchison (2009) in his classical review for a non-specialist readership.

In QFT, the physical (or quantum) vacuum is *the ground state* (a state of minimum energy) of the considered system of fundamental fields. The other states of the system are 'excited' states, containing quanta of excitation, i.e. particles. There are no particles in the vacuum (in that sense the vacuum is empty); but the vacuum is plenty of short-living virtual particle-antiparticle pairs which in permanence appear and disappear (what is allowed by time-energy uncertainty relation $\Delta E \Delta t \geq \hbar/2$).

In simple words, the quantum vacuum is a kingdom of the virtual particle-antiparticle pairs; a kingdom with apparently perfect symmetry between virtual matter and virtual antimatter.

A "virtual" pair can be converted into a real electron-positron pair only in the presence of a strong external field, which can spatially separate electrons and positrons, by pushing them in opposite directions, as it does an electric field $E$. Thus, "virtual" pairs are spatially separated and converted into real pairs by the expenditure of the external field energy. For this to become possible, the potential energy has to vary by an amount $eE\Delta l > 2m_e c^2$ in the range of about one Compton wavelength $\Delta l = \hbar/m_e c$, which leads to the conclusion that a significant pair creation occurs only in a very strong external field $E$, greater than the critical value $E_{cr} = 2m_e^2 c^3/e\hbar$, or equivalently, if the acceleration $g$ is greater than the critical value $g_{cr}$ in Equation (6).

In principle, every external force which attempts to separate particles and antiparticles, may convert a virtual pair into a real one. If it is always an attractive force, as commonly believed today, gravity can't separate particles and antiparticles. Hence, the conjectured gravitational repulsion between matter and antimatter is a necessary condition for separation of particles and antiparticles by a gravitational field and consequently for the creation of particle-antiparticle pairs from the quantum vacuum. But while an electric field can separate only charged particles, gravitation as a universal interaction might create particle-antiparticle pairs of both charged and neutral particles. Thus, the hypothesis of antigravity opens possibility for a gravitational version of the Schwinger mechanism.

The qualitative picture of the expected phenomena is very simple and beautiful. In the final stage of a hypothetical collapse, the universe would become a supermassive black hole. Deep inside the horizon of such a black hole, extremely strong gravitational field can create particle-antiparticle pairs from the physical vacuum; with the additional feature that a black hole made from matter violently repels antiparticles, while a black hole made from antimatter repels particles. Without loss of generality we may consider the case of a black hole made from matter. The amount of created (and violently repelled) antimatter is equal to decrease in the mass of black hole. Hence, during a Big Crunch, quantity of matter decreases while quantity of antimatter increases for the same amount; the final result might be conversion of nearly all matter into antimatter. If (as I will argue latter) the process of conversion is very fast, it may look as a Big Bang starting with an initial size many orders of magnitude greater than the Planck length, what may be an alternative to the inflation in Cosmology.

The most poetic part of this qualitative picture is that Big Crunch of a universe made from matter, leads to a Big Bang like birth of a new universe made from antimatter. Hence, the question why our Universe is dominated by matter has a simple and striking answer: because the previous universe was made from antimatter. There is no need to invoke CP violation as explanation for matter-antimatter asymmetry in the Universe.

For simplicity, let us consider a spherically symmetric gravitational field, created by a spherical body of radius $R_H$ and mass $M$ and let us assume that for all distances $R > R_H$, the gravitational acceleration is determined by the equation (5). This toy model allows defining a critical radius $R_{Cm}$ as the distance at which the gravitational acceleration has the critical value $g_{cr}(m)$, defined in (6). The equality $g_{cr}(m) = GM/R_{Cm}^2$, leads to:

$$R_{Cm} = \frac{1}{2}\sqrt{\bar{\lambda}_m R_S} \equiv L_P \sqrt{\frac{M}{2m}}, \tag{8}$$

where $R_S = 2GM/c^2$ is the Schwarzschild radius of a black hole with mass $M$ and $L_P = \sqrt{\hbar G/c^3}$ is the Planck length. Hence, the spherical shell with the inner radius $R_H$ and the outer radius $R_{Cm}$ should be a "factory" for creation of particle-antiparticle pairs with mass $m$. It is evident that there is a series of decreasing critical radiuses $R_{Cm}$. For instance, according to equation (8), the critical radius $R_{C\nu}$ corresponding to neutrinos is nearly four orders of magnitude larger than the critical radius $R_{Ce}$ for electrons, which is about 43 times larger than the critical radius $R_{Cn}$ for neutrons. It is obvious that if $R_H > R_{Cm}$, the creation of pairs with mass $m$ is suppressed (through the exponential factor in the equation (6)). Additionally, the equation

(8) tells us that $R_{Cm} \ll R_S$. Hence, a gravitational field, sufficiently strong to create particle-antiparticle pairs, could exist only deep inside the horizon of a black hole. An immediate consequence is that if (for instance) a black hole is made from ordinary matter, produced particles must stay confined inside the horizon, while antiparticles should be violently ejected because of the gravitational repulsion.

After integration over the volume of this spherical shell (and taking $R_{Cm} \gg R_H$), the Equation (7) leads to

$$\frac{dN_{m\bar{m}}}{dt} \approx \frac{1}{\pi}\left(\frac{R_S}{\lambdabar_m}\right)^2 \frac{c}{R_H}. \tag{9}$$

According to Equation (9), the particle-antiparticle creation rate per unit time depends on both, mass $M$ and radius $R_H$. If $R_H$ (i.e. the size of a black hole) is very small, the conversion of matter into antimatter is very fast!

Let us consider numerical examples for the critical radius determined by equation (8). With the mass of the Universe taken to be of the order of $10^{53} kg$ (see for instance Ross, 2003) the critical radius for neutrino (using $m_\nu \approx 10^{-37} kg$; according to the known bounds (Nakamura, 2010)) and neutron is respectively:

$$R_{C\nu} \sim 10^{10} m; \; R_{Cn} \sim 10^5 m = 100 km. \tag{10}$$

While we can't trust that these numbers are exact, they give an idea about the size of the universe at which the gravitational Schwinger mechanism might become important. The $R_{C\nu}$ in the equation (10) is about 10 orders of magnitude smaller than the size of our galaxy and 4 orders of magnitude smaller than the size of our Solar System.

Of course, the most interesting is to see numerical examples for the particle-antiparticle creation rate per unite time. If $R_H < R_{Cn}$ (i.e. the creation of the neutron-antineutron pairs is not suppressed), the equations (9) leads to the numerical result

$$\frac{dN_{n\bar{n}}}{dt} > 10^{86} \; pairs/s. \tag{11}$$

The numerical result (11), tells us, that decrease of matter and increase of antimatter has a rate greater than $10^{59} kg/s$, while the mass of our Universe is „only" about $10^{53} kg$ ! Such a huge conversion rate indicates that nearly the whole matter in the Universe may be transformed into antimatter (i.e. a Big Crunch of our Universe may be transformed to a Big Bang) in a fraction of second! According to this numerical example, the size of the new born Universe should be about 38 orders of magnitude greater than the Planck length, suggesting that we do not need the inflation in Cosmology.

Let us give a second, presumably extreme but instrumental numerical example, taking $R_H = 10^{-6} m$ (what is however 29 orders of magnitude greater than Planck length). If the collapsing Universe can reach such a small size, according to the Equation (8), the gravitational field is sufficiently strong to create particle-antiparticle pairs with the Planck mass $M_P$. Hence, the Equation (9) leads to the following numerical result

$$\frac{dN_{M_P \bar{M}_P}}{dt} \approx 10^{136} \, pairs/s, \qquad (12)$$

corresponding to the colossal conversion rate of $10^{128} \, kg/s$. Consequently nearly the whole matter of the Universe might be converted into antimatter in a fraction of the Planck time. In any case, the Universe is prevented to collapse to singularity; the minimal radial size of the Universe (according to apparently more realistic example (10)) might be about 40 orders of magnitude greater than the Planck length.

Hence, if there is gravitational repulsion between matter and antimatter, and if our understanding of the quantum vacuum is correct, the minimal size of our Universe should have a lower bound of the order of kilometers. According to the Inflationary Cosmology, this is a size which corresponds to the Universe *after* inflation (for friendly introductions to inflation see Linde, 2008). If the smaller sizes (and consequently higher temperatures) are not possible, we may think that an epoch of inflation and hypothesis of antigravity are incompatible.

If our scenario is correct, not only inflation, but also a number of phase transitions may not happen in the very early Universe. According to Grand Unified Theory (GUT), the primeval Universe may have developed through phases when some symmetry was exact, followed by other phases when that symmetry was broken. For instance, there are arguments that a GUT epoch (when the strong, weak and electromagnetic forces were unified) had ended with a phase transition, *before* inflation, at an energy scale of $E \approx 10^{16} \, GeV$ (Ross, 2003). Of course such a phase transition can't happen if a new cycle starts with a much lower energy.

In our opinion the key point is not if the proposed mechanism is "used" or not "used" by nature. The key point is that in principle, in the framework of quantum field theory and general relativity, the universe successively dominated by matter and antimatter is possible. It may be that we have proposed a wrong mechanism for real phenomena; if so, it stays to discover the right mechanism of such a process. In any case, it is for the first time in the eighty years after the discovery of antimatter that a cyclic universe alternatively dominated by matter and antimatter is proposed and this fascinating possibility deserves further study.

In order to avoid misunderstandings, let me underline that my conjecture concerning the gravitational proprieties of antimatter presents just one of two complementary and mutually excluding hypotheses. The other hypothesis (Noyes, 2008; Benoit-Levy and Chardin, 2009; Gilson, 2009) may be summarized as: a particle attracts both particles and antiparticles, while an antiparticle repels both particles and antiparticles. Of course, contrary to my hypotheses it is an evident violation of CPT symmetry (what can't be excluded as possibility). Such alternative postulate opens possibility that what we call dark energy is just a consequence of gravitational repulsion caused by huge quantities of antimatter located at intergalactic voids. In my approach, both, matter and antimatter are self-attracting while there is gravitational repulsion between them. The impact of antimatter is not caused by its hidden presence in the Universe but through interaction of the physical (quantum) vacuum and the ordinary matter (Hajdukovic, 2007, 2010a, 2010b, 2010c and 2011). If one of these speculations is correct it would be a quantum leap in our understanding of the Universe.